\documentclass[12pt]{article}
\usepackage{amssymb}
\usepackage{epsfig}
\def\reference{\parskip 0pt\par\noindent\hangindent 0.5 truecm}

\newcommand{\gesim}{\,\raisebox{-0.4ex}{$\stackrel{>}{\scriptstyle\sim}$}\,}
\newcommand{\lesim}{\,\raisebox{-0.4ex}{$\stackrel{<}{\scriptstyle\sim}$}\,}

\def\ga{\gesim}
\def\la{\lesim}

\def\l<{\langle}
\def\r>{\rangle}

\begin{document}
%
%
\title{The Commensal Real-time ASKAP Fast Transients (CRAFT) survey}
%

\author{Jean-Pierre Macquart$^{1,\dagger}$, M. Bailes$^2$, N.D.R. Bhat$^{2}$, G.C. Bower$^{3}$, J.D. Bunton$^4$, \and 
S. Chatterjee$^{5,6}$, T. Colegate$^1$, J.M. Cordes$^6$, L. D'Addario$^{7}$, A. Deller$^8$, \and R. Dodson$^9$, R. Fender$^{10}$, K. Haines$^9$, P. Hall$^1$, C. Harris$^9$, A. Hotan$^1$, \and S. Johnston$^4$, D.L. Jones$^{7}$, M. Keith$^4$, J.Y. Koay$^1$, T.J.W.  Lazio$^{11}$, W. Majid$^{7}$, \and T. Murphy$^5$, R. Navarro$^{7}$, C. Phillips$^4$, P. Quinn$^9$, R.A. Preston$^{7}$, B. Stansby$^1$, \and I. Stairs$^{12}$, B. Stappers$^{13}$, L. Staveley-Smith$^9$, S. Tingay$^1$, D. Thompson$^{7}$, \and W. van Straten$^2$, K. Wagstaff$^{7}$, M. Warren$^{14}$, R. Wayth$^{1}$, L. Wen$^9$ \and (the CRAFT collaboration)
}  
\date{}
\maketitle
{\center
$^1$ICRAR/Curtin Institute of Radio Astronomy, GPO Box U1987, Perth, WA 6845, Australia
\\[3mm] 
$^2$Centre for Astrophysics and Supercomputing, Swinburne University of Technology, PO Box 218, Hawthorn, VIC 3122, Australia \\[3mm]
$^{3}$Astronomy Department \& Radio Astronomy Laboratory, University of California, Berkeley, CA 94720, USA \\[3mm]
$^4$Australia Telescope National Facility, CSIRO, PO Box 76, Epping, NSW 1710 \\[3mm]
$^5$Sydney Institute for Astronomy (SIfA), School of Physics, University of Sydney, NSW 2006, Australia \\[3mm]
$^6$Astronomy Department, Cornell University, Ithaca, NY 14853, USA\\[3mm] 
$^7$Jet Propulsion Laboratory, California Institute of Technology, 4800 Oak Grove Drive, Pasadena, CA 91109, USA\\[3mm]
$^8$National Radio Astronomy Observatory (NRAO), PO Box 0, Socorro 87801 NM, USA \\[3mm] 
$^9$ICRAR/University of Western Australia, Fairway 7, Crawley, Perth, WA 6009, Australia\\[3mm]
$^{10}$School of Physics and Astronomy, University of Southampton, Highfield, Southampton SO17 1BJ, UK and Astronomical Institute `Anton Pannekoek', University of Amsterdam, Kruislaan 403, 1098 SJ Amsterdam, the Netherlands \\[3mm]
$^{11}$Naval Research Laboratory, 4555 Overlook Ave., SW, Washington, DC 20375, USA \\[3mm] 
$^{12}$Department of Physics and Astronomy, University of British Columbia, Vancouver, BC V6T 1Z1, Canada \\[3mm]
$^{13}$Jodrell Bank Centre for Astrophysics, Alan Turing Building, University of Manchester, Manchester M13 9PL, UK \\[3mm]
$^{14}$Theoretical Astrophysics Division, Los Alamos National Laboratory, Los Alamos, NM 87543, USA \\[3mm]
$\dagger$ email J.Macquart@curtin.edu.au\\[3mm]
}

%
\begin{abstract}
We are developing a purely commensal survey experiment for fast ($<5\,$s) transient radio sources.  Short-timescale transients are associated with the most energetic and brightest single events in the Universe.   Our objective is to cover the enormous volume of transients parameter space made available by ASKAP, with an unprecedented combination of sensitivity and field of view.  Fast timescale transients open new vistas on the physics of high brightness temperature emission, extreme states of matter and the physics of strong gravitational fields.  In addition, the detection of extragalactic objects affords us an entirely new and extremely sensitive probe on the huge reservoir of baryons present in the IGM. 
We outline here our approach to the considerable challenge involved in detecting fast transients, particularly the development of hardware fast enough to dedisperse and search the ASKAP data stream at or near real-time rates.  Through CRAFT, ASKAP will provide the testbed of many of the key technologies and survey modes  proposed for high time resolution science with the SKA.
\end{abstract}


\bigskip

%
%

\section{Scientific Motivation}

Short-timescale transients are associated with the highest energy density events in the Universe. They provide Nature's ultimate laboratory; their emission is usually generated by matter under extreme conditions whose properties probe physical regimes that far transcend the range achievable in terrestrial experiments.  Even the mere existence of such impulsive emission in some instances can transform our understanding of the behaviour of matter and spacetime under the most extreme conditions.  Historically, this is exemplified by the discovery of radio pulsars, and their subsequent use to test general relativity (Hulse \& Taylor 1974; Taylor \& Weisberg 1982) and the neutron star equation of state (e.g. Weber et al. 2009).

No recent event better epitomizes the problems and enormous potential scientific rewards of fast transient detection than the $\sim 30\,$Jy, 5-ms duration one-off pulse recently reported by Lorimer et al.\,(2007).  Its frequency-time of arrival sweep, if due to propagation through a dispersive medium, indicates a dispersion measure of 375\,pc\,cm$^{-3}$, which most likely places this putative object at cosmological distances.  Its extraordinary luminosity and short duration have generated a flurry of speculation as to its origin (e.g. Vachaspati 2008, Kavic et al. 2008).   If more were detected, their visibility to cosmological distances  would make them exquisite probes of the highly elusive ionized inter-{\it galactic} medium, which is the reservoir of more than half of the baryons in the present-day Universe (Nicastro, Mathur \& Elvis 2008).

One of the most astonishing implications of this interpretation of the Lorimer burst is that such events may be no rarer than GRBs but that they have only remained hitherto undetectable in the radio regime due to the lack of field of view (FoV) on telescopes with backends operating in modes capable of detecting such short events. 
This leads us to question what other events could so easily have been missed.  However, while only one  poorly localised event remains published, it is possible that the Lorimer event may not even be extraterrestrial, but may instead represent an entirely new form of atmospheric emission.  The verifiable detection of more events is thus vital, as is precise enough localization to permit optical followup capable of confirming their proposed extragalactic origin.

The key challenge hitherto impeding progress in fast transient science is thus the ability to detect transients with low event rates yet {\it simultaneously} verify and localize them to high precision.  ASKAP's marriage of phased array feed (PAF) technology with moderate-baseline interferometry overcomes this hurdle.
Viewed purely as a systematic exploration of uncharted parameter space, history suggests that  ASKAP's combination of FoV and sensitivity (Fig.\,1) make the detection of a large variety of new objects and phenomena almost an inevitability.  A high time resolution exploration of the Universe also leaves open the possibility of discovering short timescale phenomena (i.e. temporal substructure) in previously known phenomena.

Both the instrumentation we are proposing to build and the survey we intend to undertake with it represent critical steps towards the design of the SKA as a synoptic survey instrument.  There is now growing recognition that the SKA should be geared for deliberate exploration of the unknown, including the time-variable radio sky.  The importance of transients science to SKA development is underscored in the SKA science case; ``Exploration of the Unknown'' (Wilkinson et al. 2004) is a highlighted science area, and there is a a growing recognition that ``The Dynamic Radio Sky'' (Cordes, Lazio \& McLaughlin 2004) represents an anticipated payoff from an SKA design that combines widefield sampling of the sky with high sensitivity and flexibility.

Our project will operate as the proof of concept for a number of key SKA-related technologies.  In particular, it will (i) determine the suitability of PAFs in the area of transient detection; (ii) prove the use of a moderately large-N interferometric array as a cm-wavelength transient-finding machine in a field in which single dishes have historically been used for pulsar/transient detection; and (iii) it will demonstrate hardware solutions and real-time detection algorithms capable of dealing with the enormous data rates produced by next generation widefield interferometric arrays.  

Technological endeavours in the field of fast transients have historically reaped extremely large returns.
For instance, an attempt to detect radio signatures from annihilating black holes by O'Sullivan, Ekers \& Shaver (1978) is directly linked by O'Sullivan with the invention of 802.11 WLAN technology.  This is an  illustration of the extreme technical challenges posed by this field spinning off a highly lucrative development.

Even though the greatest potential of a fast transients survey lies in the completely unimagined and unknown object classes waiting to be discovered, we can already identify a large number of  guaranteed returns from the project from classes of objects already known to exist.  We describe in \S\ref{TargetsSect} some of these phenomena and how this project will contribute to our understanding of them, many of which are currently poorly understood in their own right due to their rarity. In \S\ref{PilotSect} we describe some of the preliminary surveys we are undertaking using existing instruments. These will refine the design parameters of CRAFT, particularly the regions of parameter space to concentrate on.  In \S\ref{HardwareSect} we describe the hardware and algorithm development necessitated by this technically demanding survey.  A summary of our remarks is presented in \S\ref{ConclusionsSect}.

\begin{figure}[htbp]
\centerline{\epsfig{file=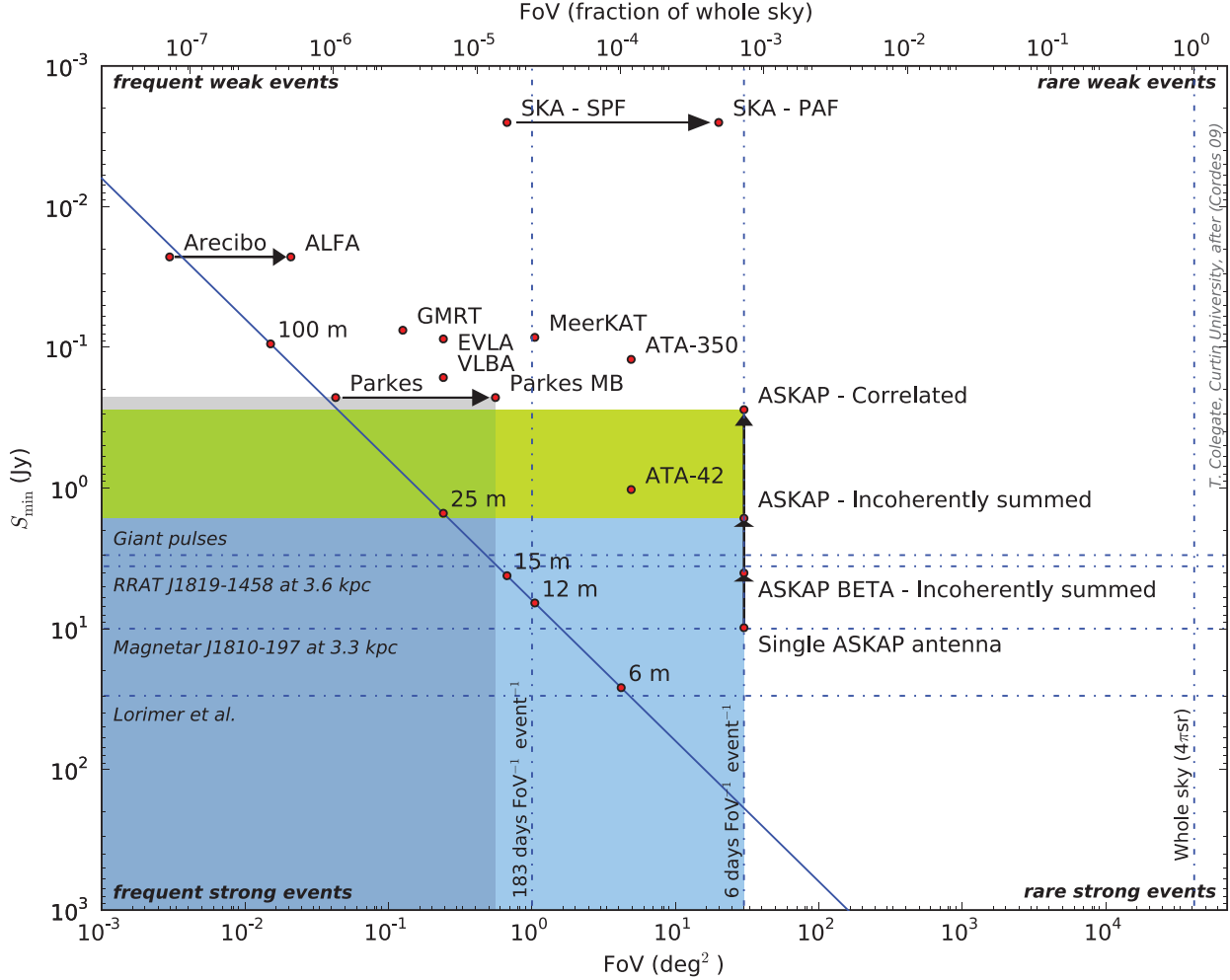,scale=0.7}}
\caption{\small Phase space probed by various current or planned radio facilities operating at 1.4\,GHz, as depicted by the minimum detectable flux density ($S_{\rm min}$) at the 5$\sigma$ level versus instantaneous accessible FoV. For simplicity in presentation, all telescopes are plotted with the same operating bandwidth (0.3\,GHz) and integration time (1\,ms). The solid diagonal line refers to single-pixel, single-reflector telescopes having --- again for simplicity --- the same system temperature (25\,K) and aperture efficiency (60\%). The SKA-SPF line indicates the sensitivity for 2000\, single-pixel feed 15-m antennas, extended to phased array feeds (SKA-PAF) with a 20 deg$^2$ FoV. The vertical dashed lines indicate the observation time required to detect an event at the rate of 225 events sky$^{-1}$ day$^{-1}$. The horizontal dashed lines show some nominal flux densities of transient sources at 1.4\,GHz. The shading represents the parameter space covered by a correlated ASKAP (green), incoherently summed ASKAP (blue) and the Parkes Multibeam (grey). Adapted from Cordes (2009).  
} \label{TelescopePhaseSpaceFig}
\end{figure}


\section{Targets}  \label{TargetsSect}

There are number of new phenomena and classes of targets potentially detectable within the auspices of a widefield radio survey for fast-timescale emission.  We discuss a selection of these, commencing with some obvious known targets, and concluding with some more speculative but high-return possibilities.

\subsection{Giant Pulses} 
All pulsars exhibit intrinsic pulse-to-pulse intensity variations, but some also emit so-called ``giant''
pulses whose strengths exceed the mean pulse intensity by several orders of magnitude. The Crab pulsar was the first found to exhibit this phenomenon.  Giant pulses with flux densities of order $10^3$\,Jy at 5 GHz and durations of $<2\,$ns, corresponding to brightness temperatures of $>10^{38}$\,K, are detected from this object (Hankins et al.\,2003). Giant pulses have since been detected from
numerous other pulsars (Cognard et al. 1996; Romani \& Johnston 2001; Johnston \& Romani
2003).  While their aggregate pulse intensities do not vary as wildly, some other pulsars exhibit a related phenomenon of ``giant micro-pulses'', where an individual component of the pulse sometimes exceeds its mean flux density by factors of $\sim 100$; the Vela pulsar is a well-studied example of this (Johnston et al. 2001).  Giant pulses may serve as probes of the local intergalactic medium (McLaughlin \& Cordes 2003 and see \S2.4).  ASKAP's contribution lies in its ability to detect giant pulses from hitherto unknown sources.  Its instantaneous FoV renders a fast transients search on ASKAP more susceptible to rare bright events than any other cm-wavelength telescope on Earth.  

\subsection{Rotating Radio Transients (RRATs)} 
The Parkes Multibeam Survey revealed a new population of sources only detectable through their individual, dispersed radio bursts (McLaughlin et al. 2006). Pulses from these sources are highly intermittent.  Although the average time between pulses ranges from 3 min to 3\,h depending on the object, fundamental pulse periods between 0.7 and 7\,s are deduced from the differences between pulse arrival times.  The resemblance of these periods to those of conventional radio pulsars strongly suggests that they are similarly associated with the rotation of neutron stars.  There are a number of outstanding questions related to RRATs.  Perhaps the most fundamental is what distinguishes RRAT from pulsar emission; this may be related to the neutron star age or magnetic field, or it may even be caused by external influences, such as the presence of a debris disk (Cordes \& Shannon 2008). Roughly only 30 RRATs are currently known.  The intermittent nature of RRAT emission renders them difficult to detect, and it is likely that the RRAT population actually rivals the number of normal pulsars.  Their discovery effectively doubles the number of neutron stars now believed to reside in the Milky Way, suggesting that the neutron star population is inconsistent with the Galactic supernova rate (Keane \& Kramer 2008).  ASKAP's large FoV makes it the ideal instrument for addressing the fundamental issue of the total number of these sources. 

\subsection{Magnetars} 
Magnetars are neutron stars whose emission is believed to be powered by decay of the object's superstrong magnetic field, rather than by the conventional mechanism of spindown.  
Recently, two magnetars have been detected at radio wavelengths with remarkably flat spectra extending to at least 100 GHz. Radio emission is episodic and bright, raising the possibility that the Galactic magnetar population may be better sampled through radio rather than X-ray surveys. 

The magnetar J1810-197 has been detected with peak single-pulse amplitudes $\sim$10 Jy and pulse widths $\sim$ 0.15 s. Given its distance of 3.3 kpc (Camilo et al. 2006), single pulses from magnetars like J1810-197 are detectable to $D_{\rm max} = D(S_{\rm pk}/S_{\rm min,SP})^{1/2} \approx 0.2$\,Mpc for Arecibo, 0.94\,Mpc for the full SKA and 70\,kpc for ASKAP. 

\subsection{Impulsive Extragalactic Transients and IGM Physics} 
Intrinsically short-duration events are susceptible to a number of effects that can be used to probe the ionized plasma through which the radiation propagates.  The two principal measurable effects are dispersion smearing and temporal broadening, due to multipath propagation.  The former measures the total column density of the plasma, while the latter is an exquisitely sensitive probe of density fluctuations within the plasma on small scales.   For instance, pulsar radiation has been used to investigate structure in the ionized interstellar medium of our Galaxy on scales from $\sim 10^{12}\,$m down to $10^5$\,m, and has provided rich insights into the physics of interstellar turbulence and the dynamics of small-scale magnetic fields pervading interstellar space.  

The detection of bright transients at extragalactic distances would open the possibility of probing the ionized Intergalactic Medium (IGM) in much the same way that pulsars have done for the ISM.   Both dispersion smearing and multipath propagation were observed in the Lorimer event, and it has been argued convincingly that neither effect is attributable to our own Galaxy's ISM.  The signal delay, if due to dispersion smearing, indicates that the event most likely occurred at $z \ga 0.2$.  

The ability of extragalactic transients to probe the IGM would strongly impact on cosmological simulations of the matter content of the universe.
Cosmological simulations indicate that the majority of baryons in the present day universe reside not in galaxies or stars, but in an extremely tenuous, $\sim 10^{-7}\,$cm$^{-3}$, IGM (Cen \& Ostriker 1999; Dav\'e et al. 2001), and that at least 45\% of these baryons have not been detected (Nicastro et al. 2008).  The missing baryons are thought to reside as highly ionized gas in a temperature range $10^5$\,K $< T < 10^7$\,K, where they are extremely difficult to detect.  Only a fraction of this ionized gas has been detected, via X-ray spectroscopy of O VII absorption lines.

Measurements of intergalactic dispersion versus redshift would enable us to investigate how the density of the IGM evolves with redshift, and the physical conditions under which most of the IGM material exists.  Do the missing baryons pervade the tenuous, warm IGM as predicted, or do they reside in otherwise undetectable partially condensed structures within it?  Does the density scale as $n_e \propto (1+z)^3$, as one expects if the IGM is still connected to the Hubble flow, or it does it already exist in structures which are partially gravitationally bound and thus disconnected from the Hubble flow? Measurements of the scattering in the IGM can be used to infer the properties of intergalactic turbulence.  Indeed, the existence of the temporal smearing in the Lorimer event may have already placed interesting limits on the amplitude of the magnetic fields pervading intergalactic space; the magnetic field needs to exceed a certain threshold value to be able to support turbulence on the small, $\sim 10^8\,$m, scales implied by the properties of the scatttered radiation (e.g. Macquart 2007).  


How many Lorimer-style bursts might ASKAP detect?  The nominal estimated rate is 225 events per day over the entire sky, equivalent to one event every 6.1 days per 30 square degree field of view.  However, this rate is based on only a single event, and is uncertain by at least half an order of magnitude.  
Detection of more such events is vital not only to confirm that these events are genuine, but also to determine their astrophysical origin.  In this regard, the log N-log S distribution based on detections of a number of these events would enable us to determine the evolution of the progenitor population throughout cosmic history.  Are their progenitors related to GRB progenitors, or do they originate from an entirely different class of phenomenon?
Precise localization of these events, made possible with ASKAP, will also permit multiwavelength followup and may provide clues as to the emission mechanism, when compared to detections (or non-detections) at higher energies.

Finally, we note that high time resolution studies of objects within our own Galaxy provide a means of studying the Milky Way's own Interstellar Medium.  However, at $\sim 1\,$ms time resolution, essentially all but the Galactic plane is accessible with little or no impact on detectability on Galactic transients from interstellar scattering at frequencies about 1\,GHz.

\subsection{Other exotica}
A variety of more speculative possibilities, discussed below, demonstrate the broad range of phenomena for which the survey will either detect, or place interesting limits on event rates.

\subsubsection{Annihilating Black Holes} 
An annihilating black hole may produce radio bursts (Rees 1977, see also O'Sullivan, Ekers \& Shaver 1978). $\gamma$-ray telescopes such as {\it Fermi} have renewed interest in the identification of high-energy signatures from primordial black holes (Dingus et al. 2002; Linton et al. 2006). Observations at $\gamma$-ray and radio wavelengths are complementary as radio observations may detect the pulse from an individual primordial black hole, while high-energy observations are sensitive to the integrated emission. 

\subsubsection{Gravitational Wave Events} 
Gravitational wave events may generate associated electromagnetic emission when matter is present in the vicinity of the system. For example, the in-spiral of a binary neutron star system may produce electromagnetic pulses at radio wavelengths due to the interaction of the magnetospheres of the neutron stars (e.g.\,Hansen \& Lyutikov 2001).  The detection of the electromagnetic signatures of events that generate gravitational waves may be necessary to localize the sources of the emission sufficiently well to uniquely identify them.  Inspiralling binary black holes act as ``standard sirens'', so that if a position and redshift can be obtained for the object, it offers a one-step measurement of the cosmic distance scale (Kocsis et al.\,2008).  It thus affords the possibility of measuring several key cosmological parameters to high accuracy, including the dark energy equation of state parameter, $w$.

\subsubsection{SETI and Extraterrestrial emitters}
Searches for extraterrestrial intelligence
(SETI), though hitherto unsucessful, have found non-repeating signals that are otherwise consistent with the expected signal from an ET transmitter.   ET signals could appear transient, even if intrinsically steady (Cordes, Lazio \& Sagan 1997).  The hardware configurations and processing techniques necessary for the detection of fast transients also satisfy most of the requirements for the detection of a SETI signal.

\subsection{Why ASKAP?}

ASKAP is poised to revolutionize the detection of fast transients for three fundamental reasons: (i) it has a large instantaneous field of view (FoV), (ii) it can localize an event to within a few arcseconds, sufficiently precise for optical spectroscopic followup, and (iii) the longer baselines in the array provide the {\it vital} aid in discriminating between genuine astronomical events and spurious terrestrial signals.

(i) A figure of merit for the detection of a population of transient objects is (Cordes 2009)
\begin{eqnarray}
{\rm FoM} = \Omega \left( \frac{A_{\rm eff}}{T_{\rm sys}} \right)^{2} K (\eta W, \tau / W),
\end{eqnarray}
with $\Omega$ the instantaneous solid angle field of view, $\tau$ the dwell time per pointing, and $K$ a function that embodies the relative detectability of transients based on their event rates, $\eta$, and event durations, $W$.  ASKAP's advantage lies in its instantaneous FoV of 30\,sq.\,deg., a factor of 49 times larger than the half-power solid angle covered by the Parkes 21-cm multibeam receiver.  The ratio of $A_{\rm eff}/T_{\rm sys}$ between Parkes and ASKAP is 1.26 (assuming a 50\,K ASKAP system temperature), giving an overall ASKAP FoM advantage of 30.

(ii)  ASKAP has the ability to localize events to within a fraction of its $\sim 7^{\prime \prime}$ synthesized beam, a much smaller diameter than that offered by Parkes, whose full width at half power is $14'$.  
Event localization is particularly important for extragalactic events, where a positional accuracy $\sim 2-5''$ is necessary to associate any putative event with a potential host galaxy and thus enable targeted optical spectroscopic followup.  For instance, positional information is crucial in confirming the proposed extragalactic origin of Lorimer-like events.

(iii) The ability to verify an event is crucial to any transient detection effort.  The history of transient detection has been dogged by the difficulty of verifying one-off events (e.g.~the Drake/Ehman WOW signal, the Lorimer burst). ASKAP's interferometric abilities, particularly the longest baselines, offer the possibility of discriminating between terrestrial and extraterrestrial events, imparting credibility to any potential transient signal detection.


%



\subsection{Survey strategy}
The CRAFT survey has adopted a purely commensal observing mode based on the fact that one patch of sky is as good as any other when surveying the sky for new phenomena whose distribution is unknown.  This is particularly true of the most extreme events which would be visible across cosmological distances, and whose sky distribution would therefore be nearly isotropic, and whose  
flux density, spectral indices, intrinsic width distributions are all unknown. 

\section{Technologies \& Pilot Surveys} \label{PilotSect}

Given that the nature of many of the objects we seek to discover is unknown, the implementation of effective hardware and survey strategies to detect them is necessarily an iterative process.  In this respect, pilot surveys are essential to evolve the project.  The evolution of the CRAFT survey will be driven by the classes of objects our pilot surveys detect (or exclude from the possibility of plausible detection), by their event rates,  the sensitivities required to investigate them, and by capitalization on advances in processing technology.

Ideally, an ASKAP fast transients detector would monitor the baseband voltage stream from all the PAFs from all the antennas.  However, the computational challenge of processing ASKAP's entire $>$12\,Tbps high-resolution data stream in real time makes this task prohibitive.  Our design strategy is therefore aimed at exploiting  the telescope's capabilities by trading FoV, temporal and spectral resolution and sensitivity against one another while maintaining a manageable data rate.  As shown in Fig.\,2, we anticipate building at least two generations of ASKAP backends, with the second generation hardware able to exploit a large fraction of ASKAP's full capabilities.

\begin{figure}[htbp]
\centerline{\epsfig{file=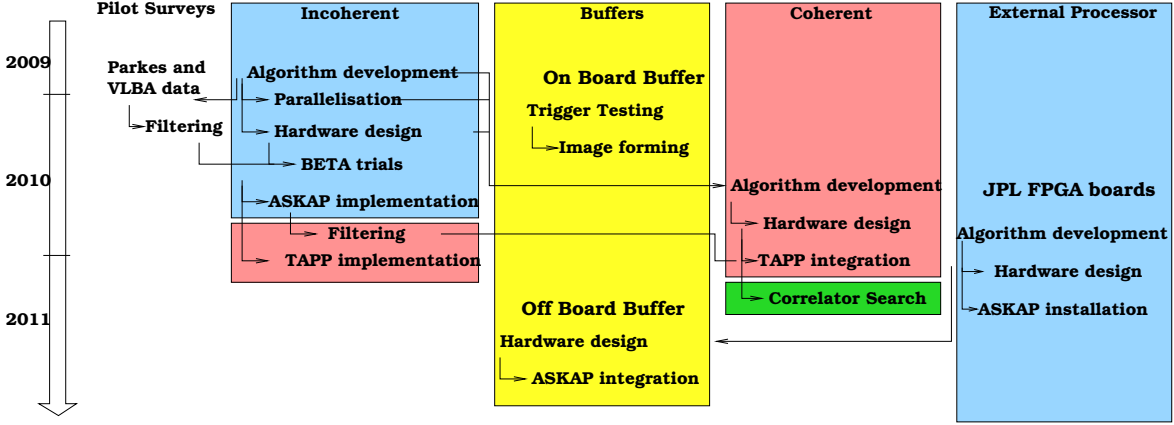,scale=0.45}}
\caption{The timeline for design studies of various fast transients machines (acronyms described in the text).} \label{TimelineFig}
\end{figure}

\subsection{Precursor surveys} \label{pilot}

Our collaboration is presently establishing a number of pilot and precursor surveys on a variety of instruments.   These are projected to all be operational within the next $\sim 6\,$months.  These surveys are being used to optimize the ASKAP survey strategy, both by establishing a number of critical estimates (i.e. restricting the range of parameter space to be searched) and proving key technologies.  In particular, they are being used to:
\begin{itemize}
\itemsep -1mm
\item {\it Establish or place upper limits on event rates as a function of flux sensitivity.}  This addresses  basic questions relating to tradeoffs between instantaneous processed FoV and sensitivity.  For instance, do we need to concentrate on searching the entire ASKAP 30\,sq.\,deg. FoV at a lower sensitivity, or is a search more effective if we restrict the searchable FoV to values $<30$sq.\,deg.\,with higher sensitivity?  
\item {\it Serve as test beds for automated transient detection algorithms.}  This is essential to any effort with ASKAP where near real-time processing and decision making is necessary (e.g. in deciding when to dump a high time-resolution, high sensitivity buffer for subsequent reprocessing).  
\item {\it Identify a means of verifying candidate events.}  This form of quality control is particularly essential in the field of transient detection, where it is typically impossible to verify a one-off event after the fact.
\item {\it Develop intelligent event characterization algorithms.}  The ability for real-time distinction of genuinely interesting events from common or uninteresting events (such as mildly bright pulses from known pulsars) is vital when processing time and/or bandwidth resources are limited.  
\item {\it Develop systems of rapidly disseminating transient event information to the broader community.}  Several collaborations, particularly in the $\gamma$-ray and optical microlensing communities, have established transient event networks to enable rapid-response followup.  Using such networks (e.g.\,VOEventNet\footnote{See {\tt http://voeventnet.caltech.edu/}}) we will publish high-confidence events and potentially also receive notification of other events which can trigger our pipelines.
\end{itemize}
Various subsets of our collaboration are conducting commensal surveys as shown in Table \ref{PilotTable}.  We describe in detail below both the VLBA and GMRT pilot surveys.  Both experiments demonstrate how existing telescopes help us to define and design fast transient experiments for ASKAP and other next generation radio telescopes.  The first stage of fast transient searches with ASKAP will likely be based on these two experiments, using the auto-correlation data from ASKAP over the much larger fields of view, in a fully commensal mode of operation.

\begin{table}[h]
\footnotesize{
\begin{tabular}{l | cc ccc c}
\null & Frequency & FoV & 1$\sigma$ Sensitivity & Time & Localization & Operating \\
\null & GHz & sq. deg. & mJy in 1\,s & resolution & radius & date \\ \hline
Parkes & 1.4 & 0.6 & 1.5 & 64$\mu$s & 14.7$^\prime$ & Sep 2009  \\
VLBA & 0.6-90 & $0.7 \,\left( \frac{\lambda}{0.21{\rm m}} \right)^2$ & 5.6 (at 0.21\,m) & 2-10\,ms & $\frac{14\,{\rm mas}}{\rm SNR} \left(\frac{\lambda}{0.21\,{\rm m}} \right)$ & Feb 2010 \\
GMRT & 0.3-0.6 & 1.2 $\left( \frac{\lambda}{0.50\,{\rm m}} \right)^2$ & 22 & 32$\mu$s & $\frac{4'}{\rm SNR} \left( \frac{\lambda}{0.50\,{\rm m}} \right)$ & end 2009 \\
LOFAR$^\dagger$ & 0.12 & 170 & 40 & $50\,\mu$s/1\,s & (3$^\prime$-10$^{\prime \prime}$)/SNR & $2010$ \\
\end{tabular}
}
\caption{Parameters of the Parkes (Swinburne), VLBA (Curtin/ICRAR), GMRT (Swinburne, Curtin/ICRAR et al.) and LOFAR (Fender, Stappers et al.) transient programs providing input to our design study. $\dagger$: time resolution is mode-dependent, localisation radius depends on whether short-only or long baselines are employed.
} \label{PilotTable} 
\end{table}

\subsubsection{VLBA pilot survey} 
A novel pilot survey for fast transients has been devised and is being implemented at the Very Long Baseline Array (VLBA)\footnote{Accepted VLBA proposal authored by Tingay, Deller and Brisken}.  
The VLBA will transition to production
correlation using the DiFX software correlator (Deller et al. 2007) before the end of 2010 for production processing of data recorded at the 10 $\times$ 25m diameter VLBA stations.  Given the flexibility of the DiFX software, the auto-correlation products for all 10 telescopes can be used for a commensal search for fast transient radio soures.  DiFX streams the auto-correlation data (averaged over short timescales of $\sim$1 ms) to a transient detection pipeline, after the data streams from all 10 telescopes have been time-aligned to high accuracy.  The transient detection pipeline can process all 10 sets of auto-correlation data separately or incoherently combine the auto-correlation data for higher sensitivity.  The pipeline will run commensally with regular VLBA correlation and whenever a candidate transient event is detected, a segment of baseband data encompassing the candidate event is extracted and
written to disk for later, more detailed analysis.

This experiment has natural advantages:
\begin{itemize}
\item The experiment is fully commensal and therefore runs almost 24/7, essentially for free, using an existing facility;
\item It will be a blind survey, in that the experiment will use pointing centres chosen for various different projects;
\item It will cover a wide range in frequency, from $\sim$1 GHz to $\sim$40 GHz, although field of view will be less at higher frequencies;
\item The use of 10 telescopes simultaneously, spanning thousands of kilometres, gives a highly robust filter against signals of a non-astronomical origin;
\item Once a candidate transient signal is detected, highly detailed follow-up processing is possible, including full coherent de-dispersion of the voltage time series, as well as potential cross-correlation of the signals (before and/or after coherent de-dispersion) in order to localise the transient with perhaps 10 mas accuracy, thereby allowing a possible identification with objects for which distances can be obtained. 
\end{itemize}

\subsubsection{GMRT fast transient survey}

An Australian-Indian collaboration (Swinburne, NCRA, Curtin/ICRAR) will conduct a survey over a small area of sky ($-10^{\circ} \leq l \leq 50^{\circ}$ and $|b| \leq 3^{\circ}$) for short-duration transients using the GMRT\footnote{Bhat et al. GMRT proposal 17\_081.}.
The survey aims to overlap with the Parkes Multibeam survey (PMB), with the region of sky deliberately chosen to have a high density of known rotating radio transient objects (RRATs). The survey will use short (5 minute) dwell times, and therefore be sensitive to objects with relatively high event rates (10 hr$^{-1}$ or more).  There are several classes of objects that the survey will be sensitive to:  giant pulse emitters, RRATs, radio magnetars and Lorimer-type bursts.

The pilot survey has both technical and scientific goals. On the technical front, the long-term aim is to develop and deliver a robust transient detection system (capable of working in commensal mode) that would become an integral part of the observatory. The expertise gained in developing such a system will contribute directly to the ASKAP fast transient system design. The survey will also generate real data capable of testing and optimising detection algorithms and observing strategies.

As with the VLBA pilot survey, the use of an interferometer for transient detection has several advantages compared to single dishes, including the ability to localise events to high accuracy, increased collecting area and an improved ability to filter RFI by comparing signals from different antennas. In addition, the GMRT survey will be the first of its kind at 610 and 327 MHz, the low frequencies also having the advantage that the beam sees a relatively large patch of sky, and that at low frequencies dispersion effects are maximised.

\section{Hardware \& Algorithm Development} \label{HardwareSect}

\subsection{Detection algorithm development} \label{algorithm}
There is no consensus on the optimal means of either identifying transients whose properties are, a priori, unknown, or of classifying the known transient types in an automated manner so as to identify those events which are particularly worthy of real-time follow up.  The latter is particularly important, as limited computing and network resources throttle the rate at which events can be processed, particularly if a large buffer must be dumped and processed following each event.

We list below the key problems in this area and the means by which they are being addressed. \\
\noindent {\it Problem 1}: Identification of genuine transients and the preliminary screening of likely events.  {\it Strategy:} we are investigating the classification of bursts in a number of dimensions of parameter space, including event duration, amplitude, sky location, dispersion measure, polarization, and recurrence rate.  
Coincidence/anti-coincidence of the signal between well-separated elements of the array are being investigated as a means of event verification and RFI rejection.  These technologies are being tested in our preliminary surveys; the effectiveness of coincidence verifications is particularly testable during the VLBA survey.  A team at Jet Propulsion Laboratory (JPL) specialising in machine learning algorithms is also investigating the deployment of their code to the preliminary transients surveys.  A number of potential algorithms/approaches, such as machine learning, have never been taken into the high time resolution astronomy domain before.   



\noindent {\it Problem 2}: Sorting the wheat from the chaff: the rejection of uninteresting transients or objects likely to otherwise trigger a transient alert, such as bright pulses from known pulsars, or outbursts from known or recurring objects. {\it Strategy:} This involves the development of a cataloguing system capable of accumulating information about known transients by position and temporal characteristics as a means of rejecting triggers on unwanted events.  The catalogue development will be guided by the precursor transients surveys, which will give us a clear indication of suitable trigger criteria.  The Swinburne, Curtin/ICRAR and NRAO team members involved in these surveys are playing a principal 
r\^ole in this development.

\noindent {\it Problem 3}: Implementation of the necessary algorithms on hardware.  {\it Strategy:} One group is investigating software solutions (Hotan, van Straten \& the Swinburne team), while another at JPL and Curtin is investigating the partial implementation of the most compute-intensive tasks onto FPGA boards.  A team at UWA/ICRAR is  developing a GPU-based search pipeline, which will be expanded to a more generalised Hough Transform-based search which makes fewer assumptions as to the signature of the dispersion signal.

\noindent {\it Problem 4}: Quantifying the relative effectiveness of various transient selection/detection approaches.  {\it Strategy:} At the conclusion of the pilot surveys there will exist a range of solutions to the detection and classifications problem.  After testing on the VLBA, GMRT and Parkes surveys, will we use the BETA system to quantify the most effective set of algorithms suited for deployment on ASKAP.

\subsection{Hardware Development: System Overview} \label{hardware}

\begin{figure}[htbp]
\centerline{\epsfig{file=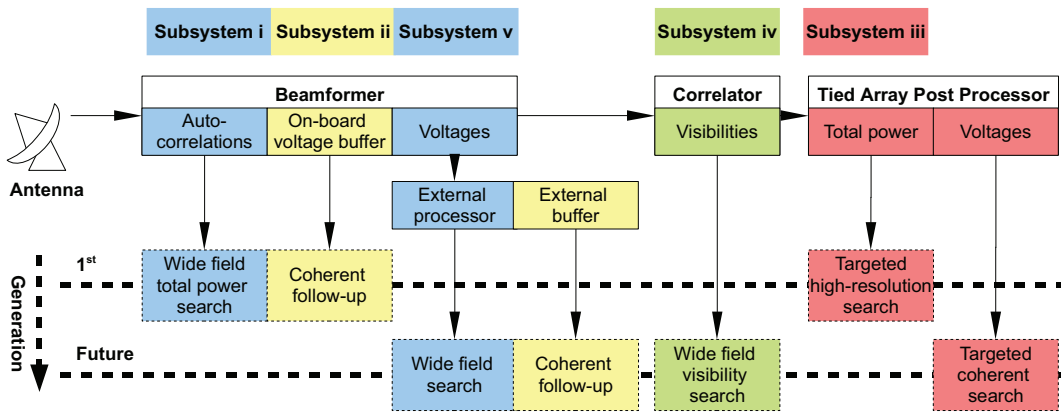,scale=0.7}}
\caption{The modular layout of our proposed detection system and its anticipated evolution with time.} \label{SysBlockDiag}
\end{figure}

We have devised an approach centred around several technologically plausible strategies aimed at exploiting various aspects of ASKAP's capabilities. These systems are scalable so as to take increasing advantage of ASKAP's full capabilities as hardware capabilities increase with time. Highlights of the aspects we are investigating are an incoherent processor, buffer implementation, advanced next-generation backends and tied-array and pipleine processing. The strategies can be broken down into subsystems, as presented in  Fig.\,\ref{SysBlockDiag} and outlined as follows: \\
{\bf i)}  an incoherent system fed from the channelised PAF beam auto-correlations; \\
{\bf ii)} a buffer for each of the PAF beam voltage datastreams. The buffer will dump the data following a trigger off another system -- nominally the incoherent search (subsystem i) -- for subsequent off-line processing; \\
{\bf iii)} a processor taking data (either voltages or powers) from a number of tied-array beams; \\
{\bf iv)} a system which will coherently search high time resolution correlator output and \\
{\bf v)} an external processor connecting to each of the PAF beam voltage datastreams.\\

\subsection{Subsystem details}
We list in this section further specifics of the various subsystems and how they are envisaged sitting within the ASKAP architecture.

\subsubsection{Data Accessed at the Beam Formers} 
\noindent {\bf First generation: Total power, incoherently summed beams (Subsystem i)} \\ \noindent
Our primary initial goal is to implement a system which monitors the 
auto-correlations, or total powers, of the incoherently summed PAF beams.
This solution exploits ASKAP's entire 30\,deg$^2$\,FoV from all
(or a subset) of its 36 antennas with the advantage of maintaining a
tractable data rate, at the expense of a loss of a factor of $\sim 6$
in instantaneous sensitivity relative to a full correlation approach. 
In the initial implementation of this subsystem, we will tap off the
auto-correlation datastreams at the post-beamformer stage, with a $\approx 1\,$ms time
resolution and $\approx 1\,$MHz spectral resolution.  Given that the initial bandwidth offsite will be 40\,Gbps or greater, the 6\,Gbps data rate from this is sufficiently low that it is feasible to transmit
this data stream off-site for real-time dedispersion and transient searching.
As part of the design study we are exploring various options for the hardware required to perform this processing. This includes GPU- and cluster-based solutions, whose compute power can be scaled with time as the data rate, complexity and scope of the transients search increases. \\
\noindent {\bf Later generation: External processor (Subsystem v)} \\ \noindent
The Curtin/ICRAR and JPL teams are forming a collaboration to investigate the construction of a next-generation fast transient detection backend. This will connect directly to the PAF beam voltage datastreams via the 10 Gbps sockets at the beamformers.  The backend is intended to provide the ultimate level of flexibility in transient searches by directly accessing the baseband (voltage) datastreams of all the PAF beams from all antennas.  This will render it capable of searching the datastream at a higher temporal and spectral resolution, and take advantage of a greater array sensitivity (i.e.\,by searching a subset of the antenna cross-powers in addition to the auto-correlations).  The specifications of the searches possible with this machine would ideally be configurable, depending on the parameters of the transients we wish to search for (e.g. high DM, temporal/spectral resolution). 
The collaboration, including an exchange of personnel, will investigate the possibility of building the backend using next-generation FPGA hardware to connect directly to the datastreams, as well as implementing dedispersion and transient search algorithms.

\subsubsection{Data Buffer (Subsystem ii)} 
A rolling buffer is envisaged as a means to access the ASKAP data stream at full sensitivity and full field of view upon demand.  Although it will initially be impossible to search the entire ASKAP data stream at full sensitivity over the entire FoV, a transient detected by another subsystem (e.g. the incoherent detection subsystem) could be used to trigger a buffer readout, which would recover a small portion of the datastream and enable us to image the event at the full array sensitivity and angular resolution, at high temporal and spectral resolution.\\
An initial buffer implementation would involve accessing the datastream from the DRAMs onboard the beamformer FPGA boards. This would provide 8GB buffers for all the data products (all telescopes, all PAFs, all polarizations). This is sufficient to retain $\sim 12$ seconds at the higher frequencies and $\sim 45$ seconds for the lower (where fewer beams are formed), allowing the full recovery of signals with DM$\la 8000$\,pc\,cm$^{-3}$.  This buffer will then be exported off-site for follow-up processing.  A later implementation, as part of the external processor, involves the connection of FPGA hardware to the 10Gbps ports of the beamformer FPGAs.  This system will hold a substantial memory buffer in its own right, relying on Moore's law to make the cost of memory cheaper in this later-generation machine.

The size of the buffer will be a crucial parameter in determining the
scope of any search.  The time to generate the trigger may be several
seconds depending on the number of trial dedispersions required and the
maximum dispersion measure, and this will impact on the amount of buffer
required and the maximum feasible search dispersion measure.  Even if 
the processing time is small, the actual event duration also effects
detection time.  For instance, an event at a DM of 4000 pc cm$^{-3}$
would be smeared to over 3.7\,s across the ASKAP bandwidth at a centre
frequency 1.4GHz.

\subsubsection{Tied-array beams (Subsystem iii)}
The tied-array beams solution represents a compromise between the characteristics of the incoherent search (with full field of view but non-optimal sensitivity) and searches using the PAF beam voltages (with full sensitivity but a prohibitive data rate for real-time continuous streaming and searching).  In effect, tied array beams allow us to access and process, at full sensitivity, at least some of the wide field of view of the PAF.  Unless we employ tied-array beams we may never be able to access the wide field of view of the PAF at full sensitivity. 

The ASKAP antenna configuration
renders it computationally prohibitive to search the entire FoV tiled  
with tied-array beams.  The diameter of a tied-array beam formed from
the entire array is only $\sim 7^{\prime\prime}$ at 1.4\,GHz.  Tied array beams are
thus most gainfully employed when either (i) a number of small
angular-size targets exists within the entire 30 sq.\,deg.~FoV or (ii) 
tied-array beams are constructed from a small number of closely-space
elements within the array, so as to increase the beam size.  Draft
specifications of the ASKAP tied-array box suggest that it will be
possible to form a maximum of 30 independent beams.

We will work towards the implementation of a tied-array system in two
stages. \\
\noindent {\bf First generation: Total power, incoherent beams} \\ \noindent
The first generation approach involves using the Tied Array Post-Processor to provide (summed) tied-array beams\footnote{In general, descriptions of beam forming in PAFs can be confusing
because there can be two stages; one for forming primary beams in the
PAFs, and one for forming tied-array beams further down the signal path.
In the ``incoherently summed approach'' of subsystem (i) (see \S4.3.1)
the total powers of each PAF beam are summed incoherently over all 36
antennas.  In subsystem (iii), which searches for
detections using tied-array beams, two stages of beam forming are
employed.  The first stage occurs at the PAF beamformers, and the second
occurs in the tied-array hardware.}
 which are subsequently squared to form total power outputs. The post-processor would ideally then integrate the signals on board (i.e.\,reduce the temporal and spectral resolution) so that the data rate is sufficiently low to enable the data to be exported off-site.  If full resolution is required the hardware will need to be on-site. \\
\noindent {\bf Second generation: Coherent, targeted, beams} \\ \noindent
Although it cannot cover the entire FoV, a tied array beam approach may be useful for tiling targeted search areas, such as the vicinity of Sgr A*, or multiple targets (e.g. individual galaxies in a nearby supercluster such as Virgo). It is also useful for SETI applications --- the requirements of our transient detection programme appears close to the requirements for SETI for targeting multiple stars/sources at high time and spectral resolution.  SETI signals would be expected to be long
lasting compared to the transients, but narrow band. This linear signature is not significantly different from that of a dispersed impulse.

\subsubsection{High Data Rate Correlator searches (Subsystem iv)}
This later-generation approach will take maximum advantage of the ASKAP processing power, i.e. the correlator, whilst bypassing the limitations imposed by the default integration time. We require access to the correlator data products on a $\approx 1\,$ms and $\approx 1\,$MHz resolution. From these baselines less than 1~km ($\sim 1/3$ of the data) will be selected to be coherently searched. This requires a search algorithm three orders of magnitude faster than that of the first generation, but will have ten times the sensitivity. A simplistic operations count implies that this should be feasible.

\subsection{Connections to other next generation facilities}

An ASKAP utilised for fast transient surveys will be a substantial
pathfinder for the next generation science to be addressed by the SKA.
 As such, ASKAP will likely be used in conjunction with other pathfinder instruments and facilities.  Possible roles for some of
these instruments and facilities are suggested here.

\subsubsection{Pawsey Centre for HPC and SKA Science}

The recent announcement of the construction of a $\$$80m high
performance computing (HPC) centre in Perth (the Pawsey Centre for HPC
and SKA Science) makes provision for substantial processing power for
ASKAP and SKA computing.  High time resolution radio astronomy is a
prime user for large amounts of computing power, given the demands of
large de-dispersion searches over massive volumes of data in real
time.  As such, CRAFT should connect strongly to the capabilities of
the Pawsey Centre as it develops on the same timescales as the ASKAP
facility.

\subsubsection{Murchison Widefield Array}

The Murchison Widefield Array (MWA) is a low frequency SKA pathfinder
currently under construction at the MRO, alongside ASKAP (Lonsdale et
al. 2009).  The MWA operates in the range 80-300\,MHz, which is complementary to ASKAP's 700-1800\,MHz frequency range.  The extremely large MWA FoV will often encompass the FoV currently being observed by ASKAP, and thus there is the potential to detect impulsive events observed with ASKAP with the MWA also.



\subsubsection{Fermi and other high energy wide field surveys}
The $\gamma$-ray telescope {\it Fermi} was launched on June 11, 2008 with a design mission lifetime of 5 years but a goal of 10 years of operations.  ASKAP will thus overlap with {\it Fermi}'s mission, even for the most pessimistic projected {\it Fermi} mission lifetime.  {\it Fermi}'s two scientific instruments, the 30\,MeV-300\,GeV Large Area Telescope (LAT) and the 150\,keV-30\,MeV Gamma-ray Burst Monitor (GBM), are both excellent transient detectors.  The LAT views $\sim 20\%$ of the sky at any one time, while the GBM  is sensitive to events over all of the sky not occulted by Earth.  There is thus considerable potential to search for the $\gamma$-ray counterparts of events detected by ASKAP.  In addition, ASKAP will be able to provide instantaneous followup of all $\gamma$-ray sources whose positions lie within the relatively large ASKAP FoV. 

\section{Conclusions} \label{ConclusionsSect}
The discovery of the fast, bright transient reported by Lorimer et al.\,(2007) has revitalized interest in short-timescale phenomena in the Universe.   It has precipitated the realization that many similar classes of objects may remain to be detected.  However, little thought has been devoted to the best means of systematically exploring the largest possible volume of this parameter space.  Over which regions of parameter space should a survey concentrate its efforts?  Should a fast transients survey concentrate on rare, bright events, or search for faint but frequent events?  Even the sorts of objects one may potentially detect is poorly constrained.    

In this paper, we have outlined a systematic approach to the exploration of this parameter space. 
We emphasise the importance of a tiered approach to the development of transients detection algorithms and hardware.  Before investing substantially in a major survey, we are employing a number of precursor surveys to either detect transients or derive significant limits on their event rates and luminosity functions.  They will foster radio transient detection algorithm development - a field very much still in its infancy - as well as indicate which regions of parameter space are likely to be most worth concentrating on.   They will also guide the specifications of the (necessarily purpose-built) survey backend hardware required.  A survey for fast transients on ASKAP will therefore represent the culmination of a long chain of efforts in searching for the optimal means of exploring this largely uncharted volume of parameter space.

Although the present paper focusses on our efforts related to ASKAP, it is obvious that the strategy applies much more widely.  A fast transients survey on ASKAP may be viewed as merely a precursor to yet more ambitious transients surveys on the SKA.

\section*{Acknowledgements}
This work was partly done at the Jet Propulsion Laboratory, California Institute of Technology, under a contract with the National Aeronautics and Aerospace Administration.

\section*{References}






\reference Camilo, F. et al. 2006, Nature, 442, 892
\reference Cognard, I., Shrauner, J.A., Taylor, J.H. \& Thorsett, S.E. 1996, ApJ, 457, L81
\reference Cordes, J.M., Lazio, T.J.W. \& McLaughlin, M.A. 2004, New Astronomy Reviews, 48, 1459
\reference Cordes, J.M. Lazio, T.J.W. \& Sagan, C. 1997, ApJ, 487, 782
\reference Cordes, J.M. \& Shannon, R.M. 2008, ApJ, 682, 1152
\reference Cordes, J.M., 2009, SKA Memo 97, Revised 04/09, http://www.skatelescope.org/pages/memos
\reference Cen, R. \& Ostriker, J.P. 1998, ApJ, 514, 1
\reference Dav\'e et al. 2001, ApJ, 552, 473
\reference Dingus, B., Laird R. \& Sinnis, G. 2002, 34th COSPAR Scientific Assembly, \#2744.
\reference Deller, A., Tingay, S.J., Bailes, M. \& West, C. 2007, PASP, 119, 318
\reference Hankins, T.H., Kern, J.S., Weatherall, J.C. \& Eilek, J.A. 2003, Nature, 422, 141 
\reference Hansen, B.M.S. \& Lyutikov, M. 2001, MNRAS, 322, 695
\reference Hulse, R.A. \& Taylor, J.H. 1975, ApJ, 195, L53
\reference Johnston, S. \& Romani, R.W. 2003, ApJ, 590, L95
\reference Johnston, S., van Straten, W., Kramer, M. \& Bailes, M. 2001, ApJ, 549, L101
\reference Kavic, M., Simonetti, J.H., Cutchin, S.E., Ellingson, S.W. \& Patterson, C.D. 2008, Jnl. Cosmology and Astroparticle Physics, 11, 17
\reference Lazio, T.J.W., Cordes, J.M., de Bruyn, A.G. \& Macquart, J.-P. 2004, New Astronomy Reviews, 48, 1439
\reference Kocsis, B., Haiman, Z. \& Menou, K. 2008, ApJ, 684, 870
\reference Linton, E.T. et al.  2006, J. Cosmol. Astropart. Phys., 1, 13
\reference Lonsdale, C. et al. 2009, Proc IEEE, 97 (8), 1497 
\reference Lorimer, D.R., Bailes, M., McLaughlin, M.A., Narkevic, D.J. \& Crawford, F. 2007, Science, 318, 777
\reference Macquart, J.-P. 2007, Proceedings of Science, Dynamic2007(022)
\reference McLaughlin, M.A. et al. 2006 Nature, 439, 817
\reference McLaughlin, M.A. \& Cordes, J.M. 2003, ApJ, 596, 982
\reference Nicastro, F., Mathur, S. \& Elvis, M. 2008, Science, 319, 55
\reference O'Sullivan, J.D., Ekers, R.D. \& Shaver, P.A. 1978, Nature, 276, 590
\reference Rees, M.J. 1977, Nature, 266, 333
\reference Romani, R.W. \& Johnston, S. 2001, ApJ, 557, L93 
\reference Taylor, J.H. \& Weisberg, J. M. 1982, ApJ, 253, 908
\reference Vachaspati, T. 2008, PRL, 101, 14
\reference Weber, F. Negreiros, R. \& Rosenfield, P. 2009 in Neutron Stars and Pulsars, Astrophysics and Space Science Library, 357, 213 (arXiv 0705.2708) 
\reference Wilkinson, P.  Kellerman, K.I., Ekers, R.D., Cordes, J.M. \& Lazio, T.J.W. 2004, New Astronomy Reviews, 48, 1551

\end{document}